%% file: main.tex
\documentclass[conference]{IEEEtran}
\IEEEoverridecommandlockouts
\usepackage{cite}
\usepackage{amsmath,amssymb,amsfonts}
\usepackage{algorithmic}
\usepackage{graphicx}
\usepackage{textcomp}
\usepackage{tabularx}
\usepackage{xcolor}
\usepackage{adp}
\usepackage{booktabs}
\usepackage{siunitx} 
\begin{document}

\title{TARS: A Theory-of-Mind Agent for Personalized In-IDE Code Comprehension}
\author{
\IEEEauthorblockN{
Leopoldo Todisco\IEEEauthorrefmark{1},
Antonio Della Porta\IEEEauthorrefmark{1},
Stefano Lambiase\IEEEauthorrefmark{2}, and
Fabio Palomba\IEEEauthorrefmark{1}
}
\IEEEauthorblockA{\IEEEauthorrefmark{1}Software Engineering (SeSa) Lab, University of Salerno, Italy}
\IEEEauthorblockA{\IEEEauthorrefmark{2}Human Augmentation and Collaboration (HAC) Group, Aalborg University, Denmark}
}
\maketitle

\begin{abstract}
Code comprehension is one of the most time-consuming tasks in software engineering, yet most LLM-based assistants produce explanations that ignore who is asking and force developers into a disruptive copy-paste workflow. We present TARS, an LLM-powered agent integrated into Visual Studio Code that supports program comprehension through autonomous explanations anchored directly to the code under analysis. Built around a lightweight Theory of Mind paradigm, TARS profiles a developer's expertise, role, and stylistic preferences, then adapts the depth and tone of its explanations accordingly, grounding them in project documentation via Retrieval-Augmented Generation. To evaluate TARS, we conducted a controlled experiment with 18 participants on non-trivial Java snippets. Participants using TARS completed tasks 26\% faster, reported lower cognitive load, and found the explanations meaningfully adapted to their profiles.

\noindent\textbf{Tool repository:} \url{https://github.com/leotodisco/tars}

\noindent\textbf{Tool video:} \url{https://www.youtube.com/watch?v=RByMKisWDLM}

\end{abstract}


\input{sections/1_introduction}
\input{sections/2_background}
\input{sections/3_tool}

\input{sections/4_evaluation}
\input{sections/5_conclusion}

\bibliographystyle{IEEEtran}
\bibliography{bibliography}

\end{document}

%% file: sections/1_introduction.tex
\section{Introduction}

Code comprehension is a fundamental activity in software engineering, underlying virtually every development task~\cite{von2002program, xia2017measuring}. Its importance is most pronounced during software maintenance and evolution, which dominate the software life cycle and consume the majority of its total cost~\cite{von2002program}, since before a developer can fix a bug, extend a feature, or refactor a module, they must first understand code they rarely authored themselves. The ability to rapidly build an accurate mental model of an unfamiliar codebase is therefore a critical, yet cognitively demanding, skill~\cite{maalej2014comprehension, ko2007information}, and developers spend a substantial portion of their time not writing code, but searching for, reading, and making sense of code written by others~\cite{ko2006exploratory, latoza2006maintaining}. This burden is compounded by the fact that internal code quality itself varies with the engineering practices adopted during development~\cite{giordano2025evidence}, leaving developers to reason about code whose structure reflects uneven practice. Among the aids that support this process, source code comments and documentation have consistently been identified as the most critical~\cite{stapleton2020human_study_of_comprehension, brooks1978using}. Yet a well-known paradox persists: despite their recognized importance, comments in practice are frequently incomplete, outdated, or incorrect~\cite{fluri2007code}, leaving insufficiently documented code as a leading source of comprehension difficulties~\cite{xia_2017_measuring}.

In recent years, researchers have begun investigating how Large Language Models (LLMs) can address this long-standing documentation dilemma\cite{della2024using}. Pioneering tools such as GILT~\cite{nam2024GILT} and IVIE~\cite{yan2024ivie} have demonstrated that LLMs can shift the developer workflow from static information \textit{retrieval} to dynamic information \textit{generation}, producing on-demand, context-aware explanations tailored to the code at hand. While these results are encouraging, existing tools remain limited in several respects. They typically target isolated code snippets and operate reactively, responding only to explicit user queries rather than proactively assisting comprehension. More fundamentally, they generate ``one-size-fits-all'' explanations that overlook the individual developer's background, skill set, and immediate goals. By failing to account for user expertise, current tools create a cognitive mismatch, either overwhelming less-experienced developers with low-level jargon or frustrating seasoned engineers with redundant, elementary details.

The difficulty of these mismatches is best understood through cognitive theories of comprehension, which cast it as an active, hypothesis-driven process: developers build understanding either bottom-up, by chunking statements into higher-level abstractions~\cite{brooks1983theory}, or top-down, by refining expectations about the code's purpose~\cite{soloway2009empirical}. Which strategy a developer follows and what they need to advance it depend on their prior knowledge or expertise, so an explanation that ignores this risks disrupting rather than supporting comprehension.

This is where Theory of Mind (ToM) becomes relevant. Originating in cognitive science, ToM describes the human ability to attribute mental states to others and to reason about what they believe, intend, and do not yet know~\cite{goldman2012tom}. It is what allows a senior developer to tailor an explanation to a colleague rather than reciting the same account to everyone. Recent work shows that this capacity can be partially elicited in LLMs through prompting~\cite{moghaddam2023boosting, wilf2024think, della2026toward, della2025unlocking, della2025prompt}, enabling assistants to maintain an implicit model of the developer's evolving understanding and shape explanations to fill gaps and avoiding cognitive overload.

To this end, we introduce \textsc{TARS}, an LLM-powered agent that enhances code comprehension through autonomous, context-sensitive assistance integrated directly into the developer's IDE. Unlike prior tools that treat LLMs as passive question-answering engines, \textsc{TARS} leverages an explicit perspective-taking mechanism to condition its explanations on the developer's expertise, role, and stylistic preferences, adapting their depth, tone, and granularity to each user. A Retrieval-Augmented Generation component further grounds explanations in project-specific documentation. By embedding the agent within the editor, \textsc{TARS} minimizes the cognitive overhead of context switching and anchors its explanations directly to the code under analysis.

We evaluate \textsc{TARS} through a controlled within-subjects experiment in which 18 participants completed code comprehension tasks on non-trivial Java snippets under both assisted and unassisted conditions. Results show that \textsc{TARS} reduces task completion time by approximately 26\%, is perceived as highly useful and easy to use, imposes significantly lower cognitive workload than the manual condition, and produces explanations that participants find meaningfully adapted to their individual profile.

%% file: sections/2_background.tex
\section{Background and Related Work}

LLMs offers a structural response to the documentation dilemma, replacing static, human-authored documentation with dynamic, on-demand explanation~\cite{nam2024GILT}. Where the traditional workflow cast the developer as an archaeologist excavating API references and forums for pre-existing answers, the emerging paradigm recasts them as a collaborator who prompts an LLM for a bespoke explanation of code in context. GILT~\cite{nam2024GILT} embodies this shift as a VS Code plugin that supports understanding through a chatbot interface, while IVIE~\cite{yan2024ivie} pushes it further by \textit{anchoring} explanations directly alongside the code they describe, mitigating the split-attention effect of toggling between the editor and a separate chat window. These tools establish two principles we adopt in \textsc{TARS}, namely that explanations should be situated within the editor, yet they share a key limitation: the explanation is the same regardless of who requests it.

Theory of Mind offers a principled way to address this. Recent work shows ToM can be elicited in LLMs through prompting~\cite{moghaddam2023boosting, wilf2024think}, and Richards and Wessel~\cite{richards2024you} brought these ideas into software engineering with TOMMY, a conversational assistant that infers a user's mental state from dialogue history to personalize explanations. \textsc{TARS} builds on this line of work but differs in two respects: it conditions explanations on an explicit developer profile rather than inferring state implicitly from conversation, and it delivers them anchored in the IDE rather than through a separate conversational channel.

%% file: sections/3_tool.tex
\section{TARS}

\subsection{What is TARS}

\textsc{TARS} is an LLM-powered assistant integrated as a Visual Studio Code (VS Code) extension that supports program comprehension directly within the developer's IDE. It is designed around two ideas: (1) explanations should be \textit{anchored} in the editor rather than delivered through a separate window, and (2) they should be \textit{personalized} for the single developer rather than identical for everyone.

To this end, \textsc{TARS} is built on a lightweight Theory of Mind paradigm. It models the developer's cognitive state, capturing their expertise, role, and stylistic preferences, and conditions its explanations on this profile to adjust their granularity, tone, and technical depth. A junior developer and a senior architect are querying the same code, therefore, receive explanations pitched at different levels. To keep its reasoning grounded in the software under analysis, \textsc{TARS} further incorporates a Retrieval-Augmented Generation (RAG) component that ingests local project documentation, enriching explanations with project-specific architectural and domain knowledge.

\subsection{How TARS works}

The explanation generation process in TARS is governed by a stateful agent built with LangGraph, structured as a directed graph where nodes encapsulate discrete processing steps and edges represent conditional transitions. The graph comprises three nodes. The \textbf{Planner Node} consumes the code snippet, the user profile, and any retrieved documentation chunks, emitting a structured JSON object with the original code and a natural language explanation. The \textbf{Syntax Check Node} validates the structural integrity of this output, looping back to the planner if malformed. The \textbf{Critique Node} performs a semantic quality assessment, triggering revision when the explanation is inaccurate or insufficient, up to a configurable retry limit.
Perspective-taking is operationalized in the Planner Node via a single consolidated system prompt that conditions the LLM's output on the profile collected by the Theory of Mind Profiler: the user's mental state, inferred through the initial questionnaire, is injected directly into the explanation prompt. While inspired by SIMTOM-Single~\cite{wilf2024think}, our approach differs in the timing and nature of the profile representation, as the cognitive profile is explicitly defined and structured \emph{before} generation, thereby reducing inferential load at execution. Rather than using multi-step inference to deduce the user's mental state at runtime~\cite{richards2024you}, TARS treats expertise level, role, and stylistic preferences as static contextual directives, optimizing latency by avoiding sequential LLM calls while retaining effective profile-conditioned adaptation. When project documentation has been indexed, the most relevant retrieved chunks are appended, grounding the explanation in the project's architecture and domain conventions.

\subsection{How to Use TARS}
Since TARS is implemented as a Visual Studio Code (VS Code) extension, the only prerequisite for installation is a working instance of VS Code.
To install the tool, the user must open the VS Code Command Palette and run

\textsc{$>$ Extensions: Install from VSIX}

\textsc{$>$ tars.vsix}

As for its use, TARS is designed to operate through three sequential steps.
First, the user must execute:

\textsc{$>$ TARS: Configure TARS}

which prompts the selection of a target Large Language Model (LLM) and the insertion of the corresponding API key. Then execute the command to access the ToM profiler.

\textsc{$>$ TARS: Theory of Mind Profiler}

\begin{figure}
    \centering
    \includegraphics[width=0.9\columnwidth]{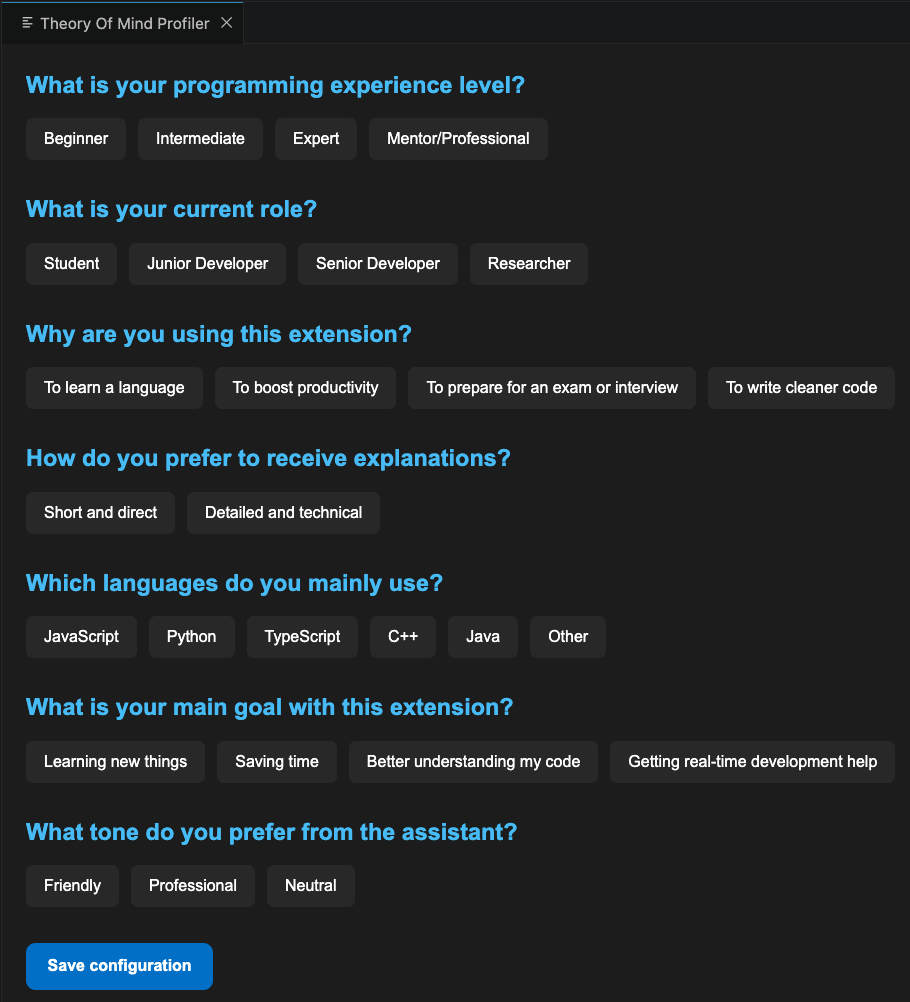}
    \caption{The Theory of Mind Profiler questionnaire presented to the user within the VSCode editor, used to capture expertise level, role, and stylistic preferences.}
    \label{fig:tars_profiler}
\end{figure}

This step presents the user with a structured questionnaire to capture their current mental state and domain expertise.
The collected profile is then used by TARS to perform perspective taking, i.e., to tailor the generated explanations to the specific cognitive and knowledge context of the individual developer.
Figure~\ref{fig:tars_profiler} shows the profiler interface as it appears within the VSCode editor.
Finally, once the profiling phase is completed, the user can request source code explanations by invoking 

\textsc{$>$ TARS: Explain Code}

\begin{figure}
    \centering
    \includegraphics[width=0.9\columnwidth]{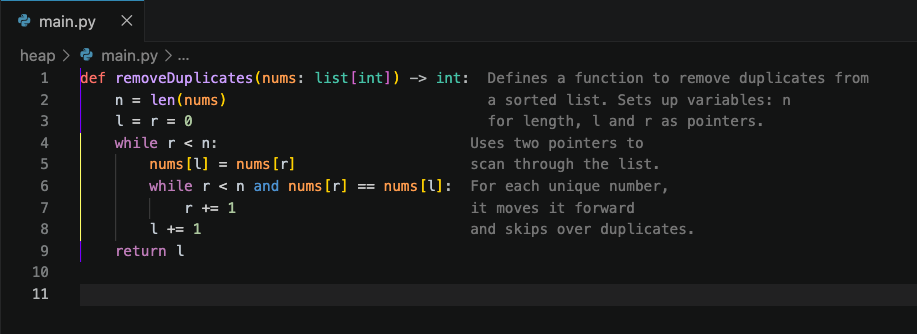}
    \caption{Example of an explanation generated by TARS for a code snippet}
    \label{fig:tars_output}
\end{figure}

The scope of the explanation \textbf{adapts dynamically to the user's selection}: if a code snippet is highlighted in the editor at the time of invocation, TARS produces a targeted explanation for that fragment; otherwise, the explanation covers the entire file currently open in the editor.
Figure~\ref{fig:tars_output} shows an example of an interaction between TARS and a developer.

%% file: sections/4_evaluation.tex
\section{Evaluation of the Tool}

\subsection{Study design}

To assess the effectiveness of TARS, we conducted a controlled experiment with 18 participants drawn from three groups: practitioners, researchers, and students. This heterogeneity is deliberate, since TARS adapts to user profiles, its effect can only be validated against users whose expertise and background genuinely differ. The overall goal of the current work is the following:

\adpbox{\faBullseye\ Goal of the work}{
Assess whether TARS effectively supports developers in understanding unfamiliar or legacy code, with respect to explanation quality, cognitive load, task performance, and adaptability to individual user characteristics.}

To reach the goal and guide our investigation, we formulated the following research questions:

\rqboxsmall{\faTasks\ \textbf{RQ\textsubscript{1}}—Does the use of TARS improve code comprehension compared to the control condition?}


\rqboxsmall{\faEye\ \textbf{RQ\textsubscript{2}}—How is TARS perceived by users in terms of utility, ease of use, and cognitive load?}


\rqboxsmall{\faSitemap\ \textbf{RQ\textsubscript{3}}—{Is TARS capable of adapting its explanations to the user's individual characteristics?}}


We employed a within-subjects experimental design to evaluate the impact of TARS on code comprehension. Under this design, all participants interacted with the same four selected Java code snippets, analyzing two snippets using the TARS-assisted condition and the remaining two without any AI assistance. 

To mitigate issues such as learning and sequence effects, we implemented a comprehensive, two-tiered counterbalancing strategy across the participants set:

\begin{itemize}
    \item \textbf{Condition Order (Macro-categories):} To prevent the results from being biased by participants simply becoming more familiar with the experimental procedure over time, the cohort was evenly divided into two macro-categories. The first group began their session using the TARS tool for the first two snippets, then switched to the manual condition for the remaining two. Conversely, the second group started with the manual control condition and subsequently transitioned to the TARS-assisted condition. This crossover approach ensures that fatigue or learning effects are symmetrically distributed across both conditions.
    
    \item \textbf{Snippet Assignment Combinations:} To ensure that the intrinsic difficulty of any specific code snippet did not skew the performance metrics of a particular condition, we systematically rotated the assignment of the four snippets. Given four snippets, there are six possible pairwise combinations for allocating them into two sets (e.g., assigning snippets 1 and 2 to TARS while assigning 3 and 4 to the manual condition; assigning 1 and 3 to TARS, and 2 and 4 to manual, etc.). Participants in both macro-categories were evenly distributed across these six combinatorial sequences.
\end{itemize}

The experiment was conducted in a controlled lab setting. Before the session begins, the first two authors explain the procedure to the participants and clarify the tasks they are expected to perform. To support them throughout the experiment, each participant receives a Markdown file that restates all the steps of the procedure. The same file specifies which scripts have to be analyzed using TARS and which ones have to be analyzed manually. To preserve the integrity of the study, participants are not allowed to use any other AI tool. The explanations produced by TARS cannot be copied and pasted, since the tool does not permit text selection, and participants are instructed not to reproduce these explanations verbatim in their responses. The first two authors remained in the laboratory for the entire duration of the experiment to ensure that every session is carried out according to these rules.

To select the snippets that users have to understand, we selected four Java code snippets sourced from the CodeSearchNet dataset~\cite{husain2019codesearchnet}, part of the CodeXGLUE benchmark~\cite{lu2021codexglue}. Snippets were selected by filtering for a cyclomatic complexity~\cite{mccabe1976complexity} between 10 and 30 using the \texttt{lizard} library, aiming for non-trivial but manageable code. All pre-existing comments were removed to avoid bias. The LLM used to run TARS was standardized to GPT-4.1 Nano across all participants, each of whom received a dedicated API key inside the task description.


To answer \textbf{RQ\textsubscript{1}}, we measured \textit{task execution time} and \textit{task correctness}. Correctness was operationalized in two complementary ways: (i) as the \textit{cosine similarity} between the participant's explanation and the ground-truth docstring, computed after a standard NLP pre-processing pipeline using the \texttt{intfloat/multilingual-e5-base} sentence-transformer model, and (ii) through a \textit{manual evaluation}, in which each explanation was independently classified as \textit{Fully Correct}, \textit{Partially Correct}, or \textit{Not Correct} against the oracle docstring by two of the authors, with a third author resolving any disagreements.


To answer \textbf{RQ\textsubscript{2}}, we use a post-experiment questionnaire that combines two validated instruments: the Technology Acceptance Model (TAM)~\cite{davis1989technology}, covering Perceived Usefulness (4 items), Perceived Ease of Use (4 items), Behavioral Intention (1 item), and Attitude Toward Use (3 items), for a total of 12 items; and a modified NASA-TLX~\cite{hart1988development}, covering Mental Demand (3 items), Physical Demand (1 item), Performance (2 items), Effort (3 items), and Frustration (3 items), for a total of 12 items, used to assess cognitive load.


To answer \textbf{RQ\textsubscript{3}}, we extended the same questionnaire described above with a dedicated custom section of 9 Trust in Machine (TOM) items, designed to assess users' perceptions of how well the generated explanations aligned with their individual profiles and backgrounds, covering dimensions such as the appropriateness of technical depth, relevance to the user's stated experience level, and overall personalization quality.

\section{Results}
We discuss the results organized by research question.

\begin{table}
    \caption{Comparison of comprehension performance under TARS-assisted and control conditions}
    \begin{tabularx}{\linewidth}{p{0.29\linewidth} p{0.18\linewidth} p{0.24\linewidth} X}
    \toprule
    \textbf{Measure} & \textbf{Condition} & \textbf{Mean (SD)} & \textbf{Median} \\
    \midrule
    Completion Time (s)  & \textsc{TARS}    & \textbf{\num{224.00}} (\num{116.74}) & \num{213.5} \\
                         & \textsc{NO TARS} & \num{304.08} (\num{215.03}) & \num{279.5} \\
                         \midrule
    Cosine Similarity    & \textsc{TARS}    & \num{0.888} (\num{0.008})   & \num{0.889} \\
                         & \textsc{NO TARS} & \num{0.886} (\num{0.007})   & \num{0.887} \\
    \bottomrule
    \end{tabularx}
    \label{tab:objective}
\end{table}

\begin{table}
    \caption{Self-reported measures (TARS condition only)}
    \begin{tabularx}{\linewidth}{p{0.40\linewidth} X X p{0.12\linewidth}}
    \toprule
    \textbf{Construct} & \textbf{Mean} & \textbf{SD} & \textbf{Scale} \\
    \midrule
    Perceived Usefulness (\textsc{PU})     & \num{4.28} & \num{0.65} & \num{1}--\num{5} \\
    Perceived Ease of Use (\textsc{PE})    & \num{4.33} & \num{0.67} & \num{1}--\num{5} \\
    Attitude Toward Use (\textsc{ATU})     & \num{4.19} & \num{0.70} & \num{1}--\num{5} \\
    Behavioral Intention (\textsc{BI})     & \num{4.61} & \num{0.61} & \num{1}--\num{5} \\
    Trust in Machine (\textsc{TOM})        & \num{3.92} & \num{0.63} & \num{1}--\num{5} \\
    \midrule
    Mental Demand (\textsc{MD})            & \num{3.64} & \num{1.56} & \num{1}--\num{5} \\
    Frustration                   & \num{3.86} & \num{1.64} & \num{1}--\num{5} \\
    Effort                        & \num{3.01} & \num{1.42} & \num{1}--\num{5} \\
    \bottomrule
    \end{tabularx}
    \label{tab:self_reported}
\end{table}

\subsection{\textbf{RQ\textsubscript{1}}—Does the use of TARS improve code comprehension compared to the control condition?}

Neither performance metric reached conventional statistical significance, yet both point in the expected direction (Table~\ref{tab:objective}). For \textit{task correctness}, cosine similarity scores were nearly identical across conditions (TARS: $M=0.888$; control: $M=0.886$), with an independent-samples $t$-test confirming no significant difference ($t(16)=-1.52$, $p=0.18$). For \textit{task execution time}, TARS participants were on average 80 seconds faster ($M=224.0\,\text{s}$ vs.\ $304.08\,\text{s}$, a 26\% reduction) and showed markedly lower variability ($SD=116.74$ vs.\ $215.03$). Because the control-group times violated normality (Shapiro–Wilk: $W=0.82$, $p<.001$), a Mann–Whitney U test was applied; the result did not reach significance ($U=494.5$, $p=0.085$), though the $p$-value lies close to the $\alpha=0.05$ threshold, suggesting a power limitation rather than the absence of a true effect. The manual evaluation corroborates this picture: under the TARS condition, explanations were more often judged \textit{Fully Correct} (7 vs.\ 4) and less often \textit{Not Correct} (2 vs.\ 3), while the majority in both conditions were rated \textit{Partially Correct}, indicating that participants generally grasped the core functionality but often missed finer details captured in the oracle docstring.

\subsection{\textbf{RQ\textsubscript{2}}—How is TARS perceived by users in terms of utility, ease of use, and cognitive load?}

All TAM constructs were rated significantly above the neutral midpoint. Perceived Usefulness ($V=2.0$, $p<.001$), Perceived Ease of Use ($V=0.0$, $p<.001$), and Behavioral Intention ($V=0.0$, $p<.001$) all exceeded the neutral value, as did Attitude Toward Using ($t=7.21$, $p<.001$). On the NASA-TLX side, Mental Demand ($t=-3.69$, $p=.0018$), Effort ($t=-5.91$, $p<.001$), and Frustration ($t=-2.94$, $p<.05$) were all significantly below the neutral midpoint, while perceived Performance was significantly above it ($V=5.0$, $p=.0003$). Participants therefore found TARS useful and easy to operate, and associated it with a low cognitive burden and high self-reported performance.

\subsection{\textbf{RQ\textsubscript{3}}—{Is TARS capable of adapting its explanations to the user's individual characteristics?}}

The mean Theory of Mind (ToM) score, calculated as the average of the scores from all ToM-related questions, was higher than the neutral midpoint of 3. This indicates that participants perceived the explanations as adapted to their individual profiles. Qualitative feedback reinforced this finding, though some participants noted a discrepancy between their declared preference for verbosity and the actual level of detail in the output, pointing to room for improvement in how the agent operationalizes the user profile.

\section{Discussions}


The evaluation of TARS provides valuable insights into the role of personalized, in-IDE AI assistants. While not all objective metrics were statistically significant, the convergence of quantitative trends and subjective feedback points to three practical implications.


\textbf{TARS as an accelerator, not a corrector.} Correctness was comparable across conditions, but TARS users were on average 26\% faster with nearly halved completion-time variance. Although this trend did not reach statistical significance, its direction suggests that TARS helps developers avoid getting stuck on unfamiliar code, making the effort required for comprehension more predictable.


\textbf{The primacy of developer experience.} The RQ\textsubscript{2} results are our clearest signal: participants significantly favored the TARS workflow, reporting lower mental demand, effort, and frustration. By offloading the burden of parsing unfamiliar syntax, TARS appears to preserve developers' mental energy, suggesting its value lies as much in reducing cognitive friction as in improving accuracy.


\textbf{The reality of AI personalization.} Participants perceived explanations as adapted to their profile rather than generic, supporting the integration of Theory of Mind into coding assistants. Yet feedback on verbosity exposes a known limitation, in that current LLMs struggle to honor abstract constraints such as ``conciseness,'' pointing to future work on stricter prompting or post-generation filtering.

%% file: sections/5_conclusion.tex
\section{Conclusions}

We presented TARS, a VS Code agent grounding autonomous, in-IDE code explanations in Theory of Mind. A within-subjects study with 18 participants showed faster comprehension, lower cognitive load, and explanations perceived as personalized. Future work will scale the evaluation to professionals on real-world codebases to test generalizability.